\documentclass[prerpint,amsmath,amssymb]{revtex4}
\usepackage{graphicx}
\usepackage{dcolumn} 
\usepackage{bm}

\def\a={&=&}
\def\beq{\begin{equation}}
\def\eeq{\end{equation}}
\def\beq{\begin{eqnarray}}
\def\eeq{\end{eqnarray}}
\def\beqa{\begin{eqnarray}}
\def\eeqa{\end{eqnarray}}
\def\bit{\begin{itemize}}
\def\eit{\end{itemize}}
\def\nnn{\nonumber \\}

\def\nnn{\nonumber \\}
\def\nnn{\nonumber \\}
\def\vr{\vec{r}}
\def\TwoFigPlace#1#2#3#4#5#6#7#8#9
{{\hbox{} \vbox to #3 true in {}
  \includegraphics{#1}
  \includegraphics{#2}
}}
\newcommand{\half}{\frac{1}{2}}   
\newcommand{\rh}[1]{\rho_{_{#1}}}

\begin{document}
\preprint{JS02}

\title{Rules for Computing Symmetry, Density and Stoichiometry 
in a Quasi-Unit-Cell Model of Quasicrystals}
\author{Hyeong-Chai Jeong}
\affiliation{
 Department of Physics, Sejong University, Seoul 143-747, Korea 
}
\author{Paul J. Steinhardt}
\affiliation{
Department of Physics, Princeton University, Princeton, 
New Jersey 08544, USA 
}


\begin{abstract}
The quasi-unit cell picture describes the atomic structure
of quasicrystals in terms of a single, repeating cluster
which overlaps neighbors according to specific overlap rules.
In this paper, we discuss the precise relationship between 
a general atomic decoration in the quasi-unit cell picture
atomic decorations in the Penrose tiling and in related
tiling pictures. Using these relations, we obtain
a simple, practical method for determining the 
density, stoichiometry and symmetry
of a quasicrystal based on the atomic
decoration of the quasi-unit cell taking proper account of
the sharing of atoms between clusters.
\end{abstract}
\pacs{}
\maketitle

\section{Introduction}
Quasicrystals are solids with quasiperiodic translational order
and rotational symmetry that is disallowed for periodic crystals,
such as ten-fold symmetry in the plane or icosahedral symmetry
in three dimensions~\cite{Levine84}.
This paper focuses on decagonal quasicrystals that
are quasiperiodic in the $xy$-plane and periodic along the $z$ 
direction~\cite{Ritsch96PML,Tsai96PML}.
The atomic structure has conventionally been described
as a periodic stacking of quasiperiodically ordered layers each of which 
can be interpreted as a Penrose tiling formed from two tile shapes. 
The two tiles correspond to two different atomic clusters which join 
to form the layer.

We have proposed an alternative picture in which the quasicrystal 
structure is described in terms of a single repeating unit called 
a ``quasi-unit cell~\cite{Steinhardt96,Jeong97,Steinhardt98Nat}''.
For ten-fold symmetric layers, the quasi-unit cell is typically
chosen to be a decagon, corresponding to a decagonal atomic 
cluster. The tiling picture and the quasi-unit cell picture are 
mathematically equivalent, but the latter has numerous advantages. 
First, the quasi-unit cell picture encodes the entire structure, 
both the symmetry and the detailed atomic decoration,
within the single decagonal 
cluster~\cite{Steinhardt98Nat,Abe00}. 
This simplifies the problem of finding
the atomic structure based on empirical data. Secondly, the 
quasi-unit cell suggests simple energetics that may explain why 
quasicrystals form~\cite{Jeong94,Steinhardt96,Jeong97}.

A key difference between quasi-unit cells and tiles is that the 
quasi-unit cells overlap. The overlap corresponds physically to the 
sharing of atoms by neighboring decagonal clusters, a feature which 
is observed in real quasicrystals~\cite{Pauling85,Burkov91,Henley91I}. 
The overlaps are restricted to certain relative positions and 
orientations of the quasi-unit cells according to ``overlap rules." 
The overlap rules replace the Penrose rules for joining tiles. 
In real quasicrystals, these overlap rules are automatically enforced 
by the atomic decoration which only permits sharing corresponding to 
the allowed overlaps. For the decagonal case, this means that the 
decoration breaks 10-fold symmetry.

The purpose of this paper is to discuss the precise relationship 
between atomic decorations in the quasi-unit cell and the rhombus 
Penrose tiling pictures, which is rather subtle, and to present a 
method for computing the density and stoichiometry of a quasicrystal 
given the atomic decoration of the quasi-unit cell. The latter 
is a simple but powerful tool for comparing
data to proposed atomic structural models. For periodic crystals, the 
analogous calculation is trivial since the density and stoichiometry
are the same as that of the unit cell because the unit cells join 
face-to-face. The technical complication for quasicrystals is finding 
a method that properly accounts for the overlap and avoids 
double-counting or under-counting. This paper presents a straightforward
method for handling the problem that has already been used
in analyzing data~\cite{Abe00}.

Section 2 discusses the relationship between the overlapping
decagon quasi-unit cell picture, the rhombus Penrose tiling and 
related tilings as ideal geometrical constructions (without
atomic decoration). Section 3 discusses the relationship between
their atomic decorations. Section 4 discusses how this relation can 
be used to construct a method for computing the density and 
stoichiometry based on the atomic decoration of a quasi-unit cell.
Section 5 discusses an interesting subclass of quasi-unit cell 
decorations, ``Almost Occupied Decoration'' (AOD) for which we have 
a very simple method of computing density and stoichiometry. 
Section 6 discusses how to classify the symmetry of the atomic 
structure based on the decoration of the quasi-unit cell.

\section{Relation between Decagon tilings and Penrose tilings}

Figure~\ref{f_overlap_rule} shows the overlap rule introduced
by P. Gummelt
which forces a perfect quasicrystalline structure with 
a single, repeating,
decagonal quasi-unit-cell, the marked decagon shown in~(a). 
The overlap rule allows two decagons to overlap only if
the shaded regions overlap and if the overlap area is 
greater than or equal to the hexagon overlap region 
of Fig.~\ref{f_overlap_rule}(b).
This permits five types of pairwise overlaps,
four types of $A$-overlaps and one type of $B$-overlap as shown
in (b). The two types of overlaps correspond to two separations
between clusters whose ratio is the golden ratio.
An infinite arrangement of decagons according to the above 
rule is called the decagonal ``covering,'' to distinguish
it from a ``tiling'' in which the units join edge-to-edge
without overlap. The covering can be mapped
into a conventional edge-to-edge tiling in at least three ways:
\bit
 \item Convert the covering to a rhombus Penrose tiling by dividing up 
central area of each decagon into a Jack (known in the Penrose 
tiling literature to 
 mean 5 fat and 2 skinny rhombi 
 surrounding a Jack vertex). This construction
 leaves three skinny rhombus-shaped regions
 inside the decagon. When neighboring decagons overlap, these regions are 
 filled in by skinny Penrose tile or a fraction of two fat Penrose tiles 
 joined together, depending on the overlap. See Fig.~\ref{f_tiling}(a).
 \item Convert the covering to a 
 three-tile ``core-area tiling'' using the core area assignment.
 The core areas correspond to an area
 assigned to a given decagon in a decagon covering
 according to whether there are $A$ or $B$ overlaps on the sides.
 The core area tiling is useful in various computations with overlapping
 decagon tilings and is a quasiperiodic tiling in its own right.
 The fundamental decagon of Fig.~\ref{f_overlap_rule}(a) is marked with
 two ``rocket''-shapes and one ``star''-shape.
 The three types of core-area tiles are determined by the overlaps
 of the star-shaped region. 
 They are a large rhombus (corresponding to two $B$-overlaps), 
 a trapezoid (a $B$-overlap on one side and an $A$-overlap on the other), 
 and a large kite (corresponding to two $A$-overlaps). 
 See Fig.~\ref{f_tiling}(b). 
 If we draw only the long edges of the core areas, we get a
 hexagon-boat-star tiling~\cite{Henley91I,Cockayne00}. 
 \item Convert to the original Penrose tiling~\cite{Penrose74} consisting
 of
 star, ship, pentagon, and rhombus tiles by joining the centers of
 decagons. See Fig.~\ref{f_tiling}(c).
 \eit
 The mapping can be reversed to produce a decagonal covering from 
 a conventional tiling.
 The equivalence between 
 the lattice structures of the decagonal covering and 
 those of Penrose tiling has been shown
 explicitly~\cite{Gummelt96,Jeong97}. 
 However, the relationship between the atomic models
 constructed by the quasi-unit-cell decorations 
 and those obtained by Penrose tile decorations
 is a different issue and 
 has not been carefully investigated previously. 
 We will call the former atomic models as 
 decagonal quasi-unit-cell models and 
 compare them with the latter Penrose-tile models.
 For Penrose-tile models, we will consider rhombus Penrose-tile
 decorations.

 \section{Atomic Decorations of Quasi-unit cells and Penrose tiles}

 We define 
 a Penrose atomic model to correspond to decorating each fat rhombus 
 identically, each skinny rhombus identically, and then joining them
 to form a Penrose tiling. A quasi-unit-cell model corresponds to 
 decorating a fundamental decagon and then covering the plane with 
 the identical decorated decagon according to 
 the overlap rule. 
 The decoration must satisfy the condition that, 
 where two atoms overlap, they are the same atom type. The rule is
 that this represents a single atom shared by the two decagon clusters.
 (There can be more than two overlaps, too, in which case one 
 atom is assigned to be shared by 3 decagons.) If an atom in one
 decagon overlaps an unoccupied site in a neighbor site, the atom is
 still assigned to the spot and shared.
 For any given site, the associated ``image'' sites are positions in
 the fundamental decagon which can overlay that site when decagons are
 joined according to the Gummelt overlap rules. The number of image
 sites varies, depending on where the original site is in the fundamental
 decagon.

 One might suppose that atomic decorations 
 of the quasi-unit-cell models form a subset of 
 the Penrose-tile models due to there being a single 
 unit and its being subject to overlap decoration constraints. 
 On the contrary, as we will show next, the set of atomic decorations of
 the decagonal quasi-unit-cell models 
 is greater than the set of atomic decorations of
 a Penrose tiling model with the 
 same edge length.

 \vspace*{.1in}
 \noindent
 {\bf Lemma I}: Every rhombus Penrose-tile model is equivalent to a 
 decagonal quasi-unit-cell model with the same edge length.

 \noindent
 Proof: A fundamental decagon can be subdivided into a Jack configuration
 of fat and skinny rhombi plus 3 skinny-shaped rhombus areas surrounding.
 Decorate each fat and skinny rhombus in the Jack according to the 
 Penrose decoration. Place no atoms in the 3 skinny-shaped 
 rhombus areas surrounding. This defines a decoration of the fundamental
 decagon. When the decagons are joined by the overlap rules, the 
 Jacks join to form a full Penrose tiling with no gaps or overlaps. 
 This result comes from the proof of equivalence between Penrose tiling
 and decagon tiling in Jeong and Steinhardt~\cite{Jeong97}.
 (The skinny outlying regions are always overlapped by Jack regions of
 neighboring decagon and thereby resolve into a skinny rhombus or parts
 of a pair of fat rhombi .)
 Hence, the Decagon decoration of the fundamental decagon obtained by this
 construction produces the same result as the Penrose
 decoration obtained by joining identically decorated
 fat and skinny Penrose tiles.

 \noindent
 {\bf Lemma II}: Some decagonal quasi-unit-cell models are not equivalent
 to rhombus Penrose-tile model with the same edge length.

 \noindent
 Proof: This Lemma asserts that the converse of Lemma I is not true.
 Consider a configuration obtained by placing one atom at the 
 center of the fundamental decagon as shown in Fig.~\ref{f_atom_deco}(a). 
 When the decagon is resolved into a Jack, one fat rhombus has an 
 atom whereas the other four do not. When the decagons are joined
 to form an overlap tiling, many obtuse rhombi remain unoccupied despite
 the overlap (see the upper part of Fig.~\ref{f_atom_deco}(a)). 
 Hence, the decagon decoration is not equivalent to a 
 Penrose-tile decoration
 based on these fat and skinny rhombi with the same edge length.

 We have shown that the atomic decorations of decagons is 
 strictly greater than the set of atomic decorations of Penrose 
 tiles of {\it identical} edge length. However, the relationship 
 between the two is more subtle than one might suppose. Let us
 next compare the atomic decorations of the decagons to decorations
 of {\it inflated} Penrose tiles whose edge length is greater
 than that of the decagons.
 For example, the quasi-unit-cell model of Fig.~\ref{f_atom_deco}(a)
 is not a Penrose-tile model with the same edge size but it is a 
 Penrose-tile model with the (single) inflated rhombi as shown in 
 the lower part of Fig.~\ref{f_atom_deco}(a).
 Each inflated fat rhombus is decorated identically and so is
 each inflated skinny rhombus. Is this true for a general 
 atomic decoration of the
 quasi-unit-cell model? The answer is no, as can be seen
 from the example of Fig.~\ref{f_atom_deco}(b). 
 The inflated fat rhombi at the right upper corner are decorated
 with zero, one or two atoms depending on the context. Neither
 doubly (at the middle of the panel) nor triply (at the bottom of the
 panel) inflated Penrose rhombi are decorated identically.
 For example, some triply inflated fat tiles have six atoms while some
 others have five. Therefore, the quasi-unit cell decoration is 
 not generally equivalent to a Penrose tiling even when one considers
 Penrose tiling edge lengths which are $\tau^3$ times that of the 
 decagon.

 However, an equivalence does emerge
 if one considers a Penrose-tile model with fourfold
 inflated Penrose tiles as shown in Fig~\ref{f_atom_deco}(c). 
 This result is surprising at first and hard to imagine
 since the fourfold inflated
 rhombus tiles, whose edge is $\tau^4$ times longer than the edge
 of the quasi-unit decagon, have scores of original decorated decagons. 
 There are nine different types of decorations of the original 
 (uninflated) decagons in general due to the nine different configurations 
 of overlapping decagons in its neighbor~\cite{Gummelt96,Jeong97} as shown 
 in Fig.~\ref{f_4th_inflated}(a). 
 In spite of this large number of different decorations, the decagons 
 at the equivalent positions in all fourfold inflated rhombus 
 tiles turn out to be the same type. Therefore, the fourfold
 inflated tiles of the same shapes are decorated identically. 
 The rigorous proof for this and the inflation rules for the
 decorated decagons will be published elsewhere~\cite{Jeong01p}.
 The right-lower panel of Fig.~\ref{f_4th_inflated}(b) illustrates our 
 claim in a finite part of decagon covering. We put the overlap-type 
 of each 
 decagon in a decagon covering and overlay them with the inflated fat 
 tile. We see that the decagon arrangement of all inflated tiles of the 
 same shapes are identical but their types are not identical for doubly
 and triply inflated tiles. However, for the quadratically inflated 
 fat tiles, we see that the types of decagons are also identical
 as shown in the right-lower panel. 

 This relationship gives a precise quantification of the degree to which
 the quasi-unit cell picture simplifies the problem of 
 defining (and finding)
 atomic structures of quasicrystals. That is, the quasi-unit cell
 description and the Penrose tiling description (allowing for 
 fourfold inflation) are equivalent mathematically, but the 
 quasi-unit cell has an area which is 
 $\frac{5}{\tau^8} \approx 0.1$
 times smaller and, hence, defines the structure using 
 $\frac{1}{10}$ times fewer atoms on average.

 \section{Comparing density and stoichiometry in quasi-unit-cell
 and tiling models}

 We define the number density of $\chi$-type atom,
 $n_\chi$ as the {\em average} number of atoms of type
 $\chi$ per unit volume in the model where 
 $\chi=\alpha, \beta, \gamma, \cdots$
 represents the atomic type of interest.
 (Greek letter, $\alpha, \beta, \gamma, \cdots$ signifies the different
 atom types.)
 For the purposes of this discussion, we will mod out the 
 periodic direction in 3D decagonal quasicrystals which can be 
 handled trivially.
 Therefore, $n_\chi$ can be considered as the 
 average number of $\chi$-type atoms per unit area in the quasicrystalline
 plane. In this plane, we take the edge of the decagonal quasi-unit-cell as
 the unit length. 

 Note that the number density of $\chi$-type atoms in any finite
 region is, in general, different from $n_\chi$ due to
 the quasiperiodicity of the quasicrystalline lattice. We define
 $n_{(\chi,C)}$ as the number density of $\chi$-type atom in 
 a finite-area $C$ for a given local context. For example,
 $n_{(\chi,C_i)}$ represents the number density
 of $\chi$-type atom in the core-area (Fig.~2(b))
 of the decagon at the center of the $i$th decagon configuration (Fig.~4(a))
 where $i=1,\ldots,9$. 
 The stoichiometry of $\chi$-type atom, $s_\chi$ is the ratio
 of the $\chi$-type atoms to the total number
 \beqa
 s_{\chi} \,=\, \frac{n_\chi}{n} 
 \,=\, \frac{n_\chi}{
 \sum \, n_\chi},
 \eeqa
 where the total number density $n$ is
 \beqa
n &=& \sum_{\chi \in \{\alpha, \beta, \gamma, \cdots\}} \, n_\chi.
\eeqa
Analogously,
the mass density $\rho_\chi$ of the $\chi$-type atom with atomic mass 
$m_{\chi}$ is 
 is
 $\rho_\chi = m_\chi n_\chi$ where 
 $\rho = \sum_{\chi} m_\chi n_\chi$ is the total mass density.

 To compute the number density in a decagon tiling,
 we can first convert to a conventional tiling, where the calculation
 is trivial because tiles join edge-to-edge.
 The only complication for tilings is for atoms at the boundaries
 where the conventional approach works. Namely, if each boundary atom is
 treated as a disk with small finite radius, the atoms at the boundaries 
 are assigned a fractional weight according to the fraction of the disk
 that lies within the tile. 

 In the previous section, we show that a quasi-unit-cell model is a rhombus
 Penrose-tile model with edge length $\tau^4$. 
 Each fourfold inflated fat ($F_4$) and skinny ($S_4$)
 tile is decorated identically. 
 Therefore, in principle,
 we can calculate the number density of $\chi$-type atom 
 $n_\chi$ in a quasi-unit-cell model in the following way. First, find the 
 decorations of $F_4$ and $S_4$ which are equivalent to the given 
 quasi-unit-cell model. Then, count the 
 number of $\chi$-type atoms in a $F_4$, $N_{(\chi,F_4)}$
 and in a $S_4$, $N_{(\chi,S_4)}$ including the 
 fractional
 atoms at the boundaries. Since the area of $F_4$ and $S_4$ are
 $A_{F_4} = \tau^8 \sin(\frac{2\pi}{5})$ and
 $A_{S_4} = \tau^8 \sin(\frac{2\pi}{10})$ respectively,
 the number density of $\chi$-type atom in $F_4$ and $S_4$ are
 given by
 \beqa
 n_{(\chi,F_4)} &=& N_{(\chi,F_4)}/ \tau^8 \sin(\frac{2\pi}{5}) \nnn
 n_{(\chi,S_4)} &=& N_{(\chi,S_4)}/ \tau^8 \sin(\frac{2\pi}{10}).
 \eeqa
 The number density of $\chi$-type atom in the model
 $n_\chi$ is now obtained by considering the fractional 
 area occupied by
 $F_4$ and $S_4$ tiles:
 \beqa
 n_\chi &=& { N_{F_4} A_{F_4} n_{(\chi,F_4)} 
 + N_{S_4} A_{S_4} n_{(\chi,S_4)}}
 \over
 { N_{F_4} A_{F_4} + N_{S_4} A_{S_4}} \nnn
 &=& {\tau^2 n_{(\chi,F_4)} + n_{(\chi,S_4)}}
 \over{\tau^2 + 1},
 \eeqa
 where $N_{F_4}$ and $N_{S_4}$ are the numbers of $F_4$ and 
 $S_4$ tiles in the tiling and we used the fact that
 both $N_{F_4}/N_{S_4}$ and $A_{F_4}/A_{S_4}$ are $\tau$. 
 
 Another way to calculate the density and stoichiometry 
 is using core areas~\cite{Jeong97}, which have the advantage that
 they are not so large. We first assign the core-tile 
 to each decagon in the decagon covering as described in 
 Fig.~\ref{f_tiling}(b) and then work out how the 
 decagon decoration translates into a decoration of core-area tiles.
 In general, the same shaped core-area tiles can be decorated differently
 due to the overlap. There are nine different decorated core-area
 tiles due to the nine different ways of surrounding a decagon of
 Fig.~\ref{f_4th_inflated}(a).
 The computational method is to determine the number density for each of 
 the nine configurations and then use the density of each 
 nine decagon-surrounding configurations in a decagon Penrose tiling. 
 Using the perp-space $r$-map volume, one can calculate
 the density of each nine configurations;

 $\begin{array}{lllll}
 \rh{1} = \tau^{-2} \rh{0},&
 \rh{2} = \tau^{-5} \rh{0},& 
 \rh{3} = \tau^{-5} \rh{0},&
 \rh{4} = \tau^{-6} \rh{0},&
 \rh{5} = \tau^{-5} \rh{0},\\
 \rh{6} = \tau^{-6} \rh{0},&
 \rh{7} = \tau^{-6} \rh{0},&
 \rh{8} = \tau^{-6} \rh{0},&
 \rh{9} = \sqrt{5} \tau^{-6} \rh{0},&
 \end{array}$ \\
 where $\rh{0} = 1/(3 \tau +1)$ and $\rh{i}$ is the density
 of the $i$th configuration of Fig.~\ref{f_4th_inflated}(a).
 If $N_{(\chi,C_i)}$ represents the number of atoms of type
 $\chi$ lying within the core-area tile of the $i$th configuration, 
 the number density for atom type $\chi$ is: 
 \beq
 n_\chi \, =\, \sum_{i=1}^{9} \,\rh{i}\, N_{(\chi,C_i)}
 \eeq
 and the ratios of $n(\chi)$ give relative stoichiometry.

\section{Simple Computation of Density and Stoichiometry} 

In the previous section, we presented two methods for calculating
the density and stoichiometry for general quasi-unit cell models.
However, both methods are somewhat cumbersome to use in practice.
To calculate the density and stoichiometry for a given quasi-unit cell 
decoration using the first method, we must figure out the atomic 
decorations of the fourfold inflated fat and skinny tiles. The second 
method  requires the atomic decoration of the 
center decagon in each of the  nine different 
configurations of  nearest-neighbor, overlapping
decagons.  A computer code is necessary to do the bookkeeping.

However, there is a subclass of decorations which includes 
the cases of most practical interest which can be calculated by hand.
This class, which we call the  ``almost occupied decoration (AOD),''
encompasses all decorations
where core areas are decorated the same for any of
the nine-configurations, so we can forego treating each of the 
nine separately. This class includes most decorations with 
physically reasonable densities. It is still includes and is
larger than the Penrose tile decorations (with same edge length).

First, let's consider the even simpler
``fully occupied decoration (FOD)'' in which overlap does
not add any atoms to the core areas:

\noindent
{\bf Definition:} fully occupied decoration (FOD) --- A single decagon
decoration is a fully occupied decoration if, for every atom in the 
decagon, the image sites {\it lying within the ``big kite'' core area} 
are all occupied by the same atom type. 

Here, the ``big kite'' core area  is the kite core area shown in 
Fig.~\ref{f-region}(a). Note that images outside the big kite core 
area may be unoccupied. The special property of FOD is that overlaps
can be ignored in computing stoichiometry.  The big kite core area is the  
largest of the core areas and all other core areas are subregions.
If the decoration is fully occupied, then {\it every} atom anywhere in
the decagon has all occupied image sites inside the big kite (and, 
therefore, inside each of the core areas).  Therefore, overlapping 
neighbor tiles does not add any new atoms inside core areas in any case.
Therefore, the stoichiometry can be computed by considering only three
core-area types.

Let three Roman letters $A$, $B$, $C$ will signify the three types of 
decagons: 
\begin{itemize}
\item $A =$ decagons with two $A$-overlaps, 
\item $B =$ decagons with two $B$-overlaps, 
\item $C =$ decagons with one $A$- and one $B$-overlap.
\end{itemize}
Here, we refer only to overlaps on the two sides of decagon. 
(In Fig~\ref{f-9confOL}, the center decagons in the 
configurations 1 and 2 are A type, 3 -- 6 are C type and 7 -- 9 
are B type.) 
Associated with each type of decagon above, we can assign 
a ``core area'' of the decagon.
\begin{itemize}
\item $A$: kite shape
\item $B$: rhombus shape
\item $C$: trapezoid shape
\end{itemize}
as shown in Fig.~\ref{f_tiling}(b) and these core areas join to 
completely cover the plane with no gaps.

Let $N_{(\chi,X)}$ represent the number of atoms of type 
$\chi$ lying within the core area of an $X$ decagon
where $X = A, B$, and $C$. Note that $N_{(\chi,X)}$ can be counted
directly from the fundamental decoration (the decagon decoration before
overlap) since overlaps with the neighbor decagons do not add any 
new atoms inside core areas. The number density $n_\chi$ for atom type 
$\chi$ is now given by
\beqa 
 n_\chi	\a= \rh{A} N_{(\chi,A)} + \rh{B} N_{(\chi,B)} 
	+   \rh{C} N_{(\chi,C)}, 
\label{e-nchi}
\eeqa
where,
$\rh{A} = 2 \tau^{-3} \rh{0}$,
$\rh{B} = \tau^{-3} \rh{0}$, 
and $\rh{C} = 2 \tau^{-4} \rh{0}$.

Equivalently, we can divide the fundamental decagon 
into regions and assign the weight for each region
as shown in Fig.~\ref{f-region}(b). 
Since the the core area $A$ (kite shape) 
and $C$ (trapezoid) include the core area $B$ (rhombus shape), 
the atoms in the region~I should be counted fully
when we calculate the number density.
Region~II does not belong to
the core area $B$ but 50 percent of it
belongs to the core area $C$ 
and the whole area belongs to core area $A$.
Therefore, the number density weights, $\rh{I}$ and $\rh{II}$
assigned to  regions~I and~II are given by
\beqa
 \rh{I} \a= \rh{A} + \rh{B} + \rh{C} = \rh{0}			\nnn
 \rh{II} \a= \rh{A} + \half \rh{C}   = \tau^{-1} \rh{0}.
\eeqa

Now, we introduce the AOD class models, for which
the same shape of core areas are decorated identically 
as for the FOD class models
but the core area decoration (after overlap)  may be different 
from that of the fundamental decoration (before overlap).

\vspace*{2mm}\noindent
{\bf Definition:} almost occupied decoration (AOD) 
--- A single decagon decoration is an almost occupied decoration 
if, for every atom in the decagon {\em outside of the  ``tiny central kite''
area shown in Fig.~\ref{f-region}(c)}, the image sites lying within
the ``big kite" core area are all occupied by the same 
atom type. 

\vspace*{2mm}\noindent
{\bf Lemma IV}: All core areas of the same shape have the 
same decoration after overlap.

This Lemma means that, as with the FOD, we do not have to consider
each of the nine local arrangements of decagons separately.
\vspace*{2mm}\noindent 
Proof: For an FOD class model, the same shape of 
core areas have the same decoration since the overlap
does not bring any new image atoms to the core areas.
For an AOD class model, new image atoms may exist
in the tiny central kite area. However, due to the way the 
decagons overlap, 
the image atom arrangements in the central kite 
are the same for each type of  core, areas as shown 
in Fig.~\ref{f-9confOL}. 
When the central core area is the big kite shape (Conf.-1 and~2), 
overlapping decagons cannot add any images on the central kite 
area. Therefore, the big kite-shaped core areas (for both 
Conf.~1 and~2) have the same images.
When the central core area is the rhombus shape (Conf.-7,~8 and~9), 
two neighboring decagons overlap with the tiny central kite
area. For all three configurations, the relative position and 
the direction of the overlapping decagons are the same. Therefore, 
the image sites in the tiny central kite area are the same 
for all three cases. Finally,
 When the central core area is the 
trapezoid shape (Conf.-3, 4, 5, and 6), only one neighboring 
decagon can overlap the tiny central kite area. The position 
and the direction of the overlapping decagons are the same for
the same orientation of the trapezoid (Conf.-3 and~4, and 
Conf.-5 and~6 are the same orientation respectively).
Therefore, the trapezoid-shaped core areas with the same orientation 
have the same images. Altogether, the image sites in the tiny
central kite area is the same decoration for the same shape of
core areas.  

\vspace*{2mm}
Since the same shaped core areas are decorated identically,
the number density $n_\chi$ for an AOD model can be
calculated from Eq.~(\ref{e-nchi}) provided that
$N_{(\chi,X)}$ is the number of atoms in the the core area X
($X=A,B,C$) {\em including} image atoms. We can divide
the fundamental decagon into regions with the weight
for each region for AOD case too as shown in Fig.~\ref{f-region}(d).
Here, the atoms in the region~III may bring new image atoms 
in the tiny central kite area and hence must be taken into account
for the number density of the atoms. As mentioned,
region~III does not add any image atoms to the 
tiny central kite area for the A-decagons (Conf.-1 and~2) 
but always adds images for the B-decagons (Conf.-7,~8, and~9)
and 50 percent for the C-decagons (Conf.-3,~4,~5 and~6).
Therefore, the number density weights are given by 
\beqa
 \rh{I} 	\a= \rh{0}		\nnn
 \rh{II} 	\a= \tau^{-1} \rh{0}	\nnn
 \rh{III} 	\a= \rh{B} + \half \rh{C} = \tau^{-2} \rh{0}
\eeqa
for the AOD class models. 
Note that the overlap does not add the new image atom if 
there was an atom at the image site in the tiny central 
kite area of the fundamental decoration. Therefore, the weight
$\rh{III}$ should be assign to the atoms in the region~III
only if their images sites (in the tiny central area) are unoccupied.
Also, it is possible that the image 
sites from the left and the right part of  region~III coincide
as in the case of the mirror symmetric decoration. 
For these atoms, we should count them only once for the type-B decagons 
and, therefore, we should assign 
\beqa
 \rh{III}' 	\a=  \half (\rh{B} + \rh{C})	\nnn
		\a=  (\half \tau^{-3} + \tau^{-4}) \rh{0}
\eeqa
to the atoms in the regions~III 
(whose image sites in the tiny central area are unoccupied).

While the description above may seem complex, a brief study will
show that it is straightforward to apply the method using the 
back of an envelope for a wide 
class of useful decorations.

\section{Symmetry of a Quasi-Unit-Cell model}

 The unit cell construction in periodic crystals is used not only
 to determine the density and stoichiometry, but also to classify
 the symmetry. In this section,
 we analyze the symmetry of quasi-unit-cell models for decagonal
 quasicrystals whose atomic sites can be interpreted as projections
 from the higher dimensional hypercubic lattices. 
 In general, the atomic 
 positions in the fundamental decagon can be in 
 any location as long as they 
 satisfy the overlap decoration rule. However, here we only consider
 the fundamental decagon decorations whose atomic positions, 
 relative to the position of a vertex of the decagon, are given by
 \beqa
 \vr &=& \sum_{k=0}^{4} m_k \hat{e}_{k},
 \eeqa
where $m_k$ is an integer and 
$
\hat{e}_{k} = \cos(\frac{2k\pi}{5}) \hat{x} + \sin(\frac{2k\pi}{5}) \hat{y}
$
is a unit vector in the physical space. 
Mathematically speaking, this quasilattice of
points is a set of measure zero compared to the entire 2d interior
of the decagon, 
 but this does not impose any practical restriction
 on constructing physical atomic models since the quasilattice
 points are dense 
 everywhere due to the irrationality between the basic vectors 
 $\hat{e}_{k}$s.

 For these models, atoms can be lifted to the five dimensional 
 hypercubic lattice points and the information on the correlations 
 between positions of atoms in the model can be encoded into the geometry 
 of the perp-space images which is the set of points given by
 \beqa
 \vr^{\perp} &=& \sum_{k=0}^{4} m_k \hat{e}^{\perp}_{k},
 \eeqa 
 where
 $\hat{e}^{\perp}_{k} = 
 \cos(\frac{4k\pi}{5}) \hat{x}^{\perp}
 +\sin(\frac{4k\pi}{5}) \hat{y}^{\perp}
 $
 is a unit vector in a plane perpendicular to the physical space. 
 When the perp-space images of two models 
 have the same geometrical shape, they are
 indistinguishable~\cite{Mermin91I}
 since the correlations between atomic positions 
 are fully determined by the shape of the perp-space image.
 Following Mermin~\cite{Mermin91I}, we call an operation ${\bf g}$
 a symmetry operation if the state after the operation is 
 indistinguishable from the state before the operation. 
 That is, if ${\bf g}$ is a symmetry operation for a model with
 density function $\rho(\vr)$ then $\rho(\vr)$ and 
 $\rho'(\vr) := \rho({\bf g}^{-1}\vr)$ should have the 
 identical correlation functions.
 Therefore, an operation ${\bf g}$ represented by 
 \beqa
 {\bf g} \hat{e}_k &=& \sum_{j} g_{jk} \hat{e}_j 
 \eeqa 
 is a symmetry operation if the corresponding perp-space
 operation ${\bf g^\perp}$ given by 
 \beqa
 {\bf g^\perp} \hat{e}^{\perp}_k &=& \sum_{j} g_{jk} \hat{e}^\perp_j.
 \eeqa 
 preserves the geometry of perp-space images. 

 As one can see from the definition of $\hat{e}_k$ and
 $\hat{e}^\perp_k$, $\frac{2\pi}{10}$ rotation in the real space
 corresponds to $\frac{4\pi}{10}$ rotation in the perp-space
 since it is equivalent to $\hat{e}_k\rightarrow \hat{e}_{k'}$
 with $k' = k+1$(mod 5).
 A tenth of $2\pi$ ($\frac{\pi}{10}$) rotation in the real space is 
 equivalent to
 $\hat{e}_k\rightarrow -\hat{e}_{k'}$
 with $k' = k+3$ (mod 5) and therefore corresponds to
 $\frac{7\pi}{10}$ rotation in the perp-space. 
 Since the $\frac{7\pi}{10}$ rotation (with the modulus $2\pi$) 
 generates all ten integer multiples of $\frac{\pi}{10}$, 
 the quasi-unit cell model has ten-fold symmetry if and only 
 if the perp-space image has ten-fold symmetry. 
 Also, the quasi-unit cell model has reflection symmetry to an $\hat{e}_n$-axis if 
and only if the perp-space image has reflection symmetry to the $\hat{e}^\perp_n$-
axis.
 This is because the the reflection to the $\hat{e}_n$-axis in 
 real space corresponds to the reflection to the $\hat{e}^\perp_n$-axis 
 in the perp-space. They are equivalent to 
 $\hat{e}_k\rightarrow \hat{e}_{k'}$ 
 or
 $\hat{e}^\perp_k \rightarrow \hat{e}^\perp_{k'}$ 
 with $k+k' = 2n$ (mod 5) 
 in both real and perp spaces. 

The geometry of the perp-space image for a quasi-unit-model can be 
obtained from the perp-space image of the atoms in the fundamental decagon. 
Let us first consider a simple case where the fundamental decagon is 
decorated with only one atom. In this case, the perp-space image of the 
atoms in the decagons with a certain direction is a simply connected 
triangle shape. 
(The Gummelt decoration of a decagon breaks the 10-fold symmetry of the
 decagon. We can define the ``direction'' of a (Gummelt decorated) decagon
 from this. In Fig.~\ref{f_deco_Jack}, we define the direction of a 
decagon along the 
 mirror symmetry line, from the intersection of two ``rockets'' to 
 the center of the ``star''. The decagon in a Gummelt covering 
 has 10 different possible directions and the perp-space image of the 
 Jack vertices in the decagons with the same direction is the triangle 
 shown in Fig~\ref{f_deco_Jack}(b).)
The perp-space images of all atoms in the model is then given by 
 the union of ten triangles of ten different directions.
 Figure~\ref{f_deco_Jack}(a) shows such simple decoration of fundamental 
 decagon 
 which is equivalent to a Penrose lattice decoration where atoms 
 decorate each Jack vertex. The perp-space image of this 
 quasi-unit-cell model is simply the perp-space occupation domain of 
 Jack vertices~\cite{Bruijn81b} and given by Fig.~\ref{f_deco_Jack}(b). 
 Here $J_{k}$ and $J_{\bar{k}}$ represent the perp-space image of Jack 
 vertices in the $\hat{e}_k$ direction and the $-\hat{e}_k$ direction 
 respectively. Since the perp-space image has 10-fold symmetry and
 the mirror symmetry, the quasi-unit-cell model with the fundamental 
 decoration of Fig.~\ref{f_deco_Jack}(a) has $D_{10}$ symmetry.

 The symmetry for a general fundamental decoration can be obtained
 similarly. If the relative position (from a Jack vertex)
 of a decorated atom is given by $\vr = \sum_{k=0}^{4} m_k \hat{e}_{k}$,
 then the perp-space image from the decagons with the 
 $\hat{e}_0$ direction
 will be the triangle given by the 
 $\vr^\perp = \sum_{k=0}^{4} m_k \hat{e}^\perp_{k}$
 translation of the $J_0$ triangle.
 The perp-space image from the decagons with the $\pm\hat{e}_k$ direction
 will be obtained
 by rotating the $\hat{e}^\perp_0$ direction triangle by the angle that 
 the $\pm \hat{e}^\perp_k$ and the $\hat{e}^\perp_0$ axes make.
 The perp-space image of all decagons is then the union of these 
 10 triangles and always has 10-fold symmetry.
 Therefore we can conclude that all quasi-unit-cell models
 have 10-fold symmetry. 

 How about the mirror symmetry? The quasi-unit-cell model with 
 the fundamental decoration of Fig.~\ref{f_deco_Jack}(b) has the 
 the mirror symmetry as well as the 10-fold symmetry.
 Is the mirror symmetry also general for quasi-unit-cell models?
 The answer is no, as one can see from the example of 
 Fig.~\ref{f_deco_no_m}(b). In this example, the perp-space 
 image has no mirror symmetry axis. 
 Then when does the perp-space image has the mirror symmetry?
 From the fact that the reflection transforms the base vectors in 
 the perp-space equivalently to the transform of base vectors 
 in the real space, we can say that the perp-space image has the 
 mirror symmetry if the decoration of fundamental decagon has the 
 mirror symmetry. However, the converse is not true. Some asymmetric 
 decorations of the fundamental decagon can produce a quasi-unit-cell model
 with the mirror symmetry as shown in Fig.~\ref{f_deco_no_m}(c) 
 and~(d). 
 
 In sum, a quasi-unit-cell model has the 10-fold symmetry.
 If the decoration of fundamental decagon has a mirror symmetry,
 then the quasi-unit-cell model has a mirror symmetry as well as 
 the 10-fold symmetry.

 \section{Concluding Remarks}
 We have shown that a Penrose-tile model is a quasi-unit-cell model
 with the same edge size and a decagonal quasi-unit-cell model is a 
 Penrose-tile model with the quadratically inflated rhombi. 
 This means the set of all decagonal quasi-unit-cell models
 is the same as the set of all rhombus Penrose-tile models.
 However, mathematical equivalence between the two sets
 does not imply that they are physically equivalent in constructing 
 atomic models for quasicrystals. 
 We need only one kind of atomic clusters as a building 
 unit in quasi-unit-cell models while at least two basic 
 building units are needed in Penrose-tile models.
 Furthermore, the building unit size is much smaller for 
 quasi-unit-cell models. For a general quasi-unit-cell decoration,
 the corresponding Penrose tiles, whose decorations are identical, 
 have scores of original quasi-unit-cell. 
 In other words, we have to look for more and larger
 building units when we use Penrose tile models.

This work was supported by Korea Research Foundation Grant 
KRF-2001-041-D00062 (HCJ) and US Department of Energy grant
DE-FG02-91ER40671 (PJS).


\clearpage

 \section*{Figure Captions}
\begin{figure}[!h]
 \includegraphics[width=90mm]{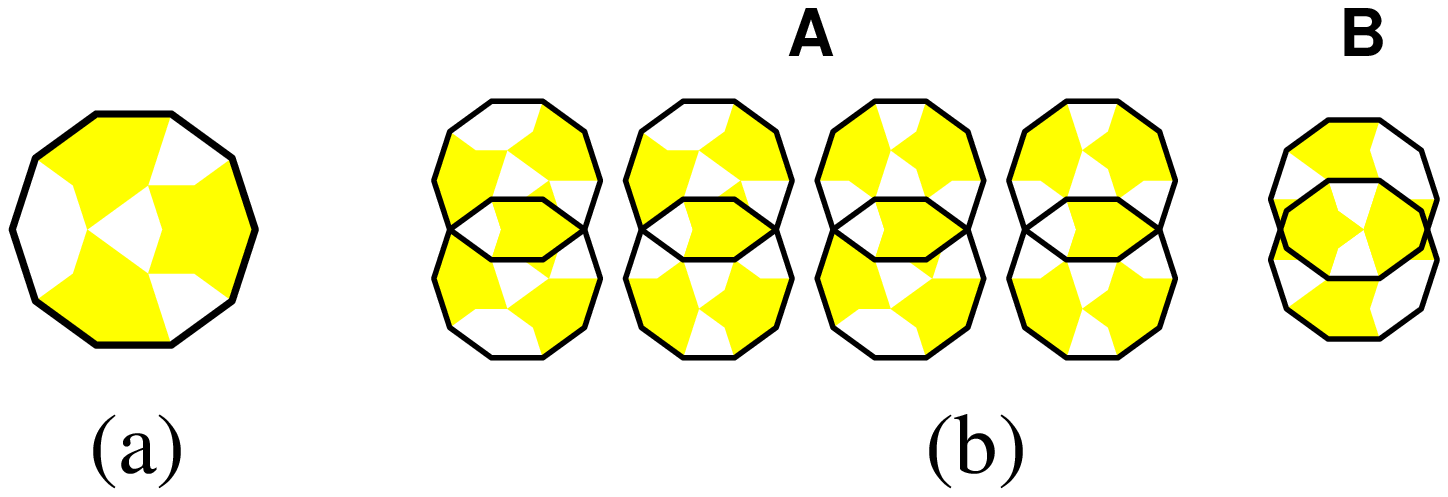} 
 \caption{
 }
 \label{f_overlap_rule}
 \end{figure}

\begin{figure}[!h]
 \includegraphics[width=90mm]{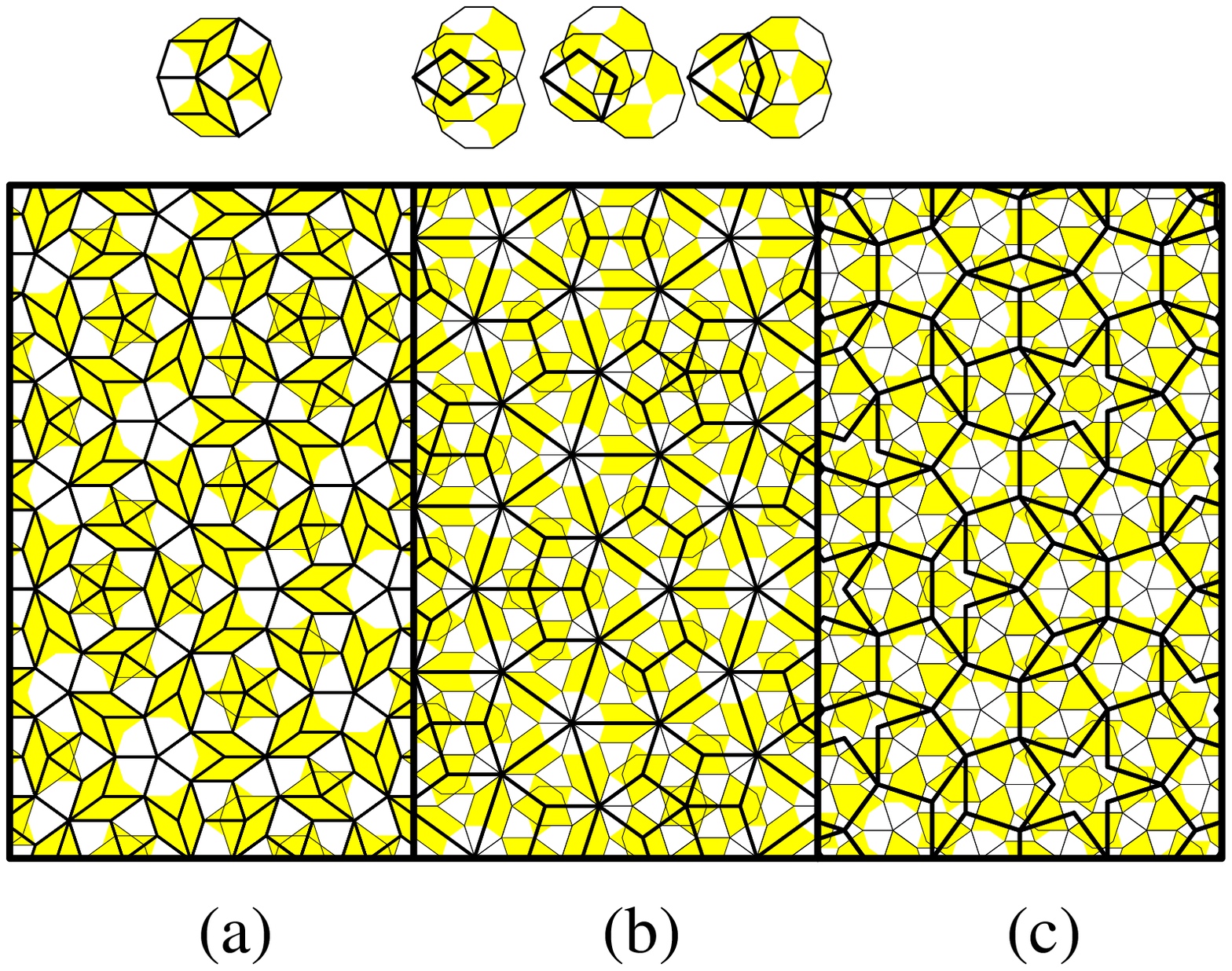} 
 \caption{
 }
 \label{f_tiling}
 \end{figure}

\begin{figure}[!h]
 \includegraphics[width=90mm]{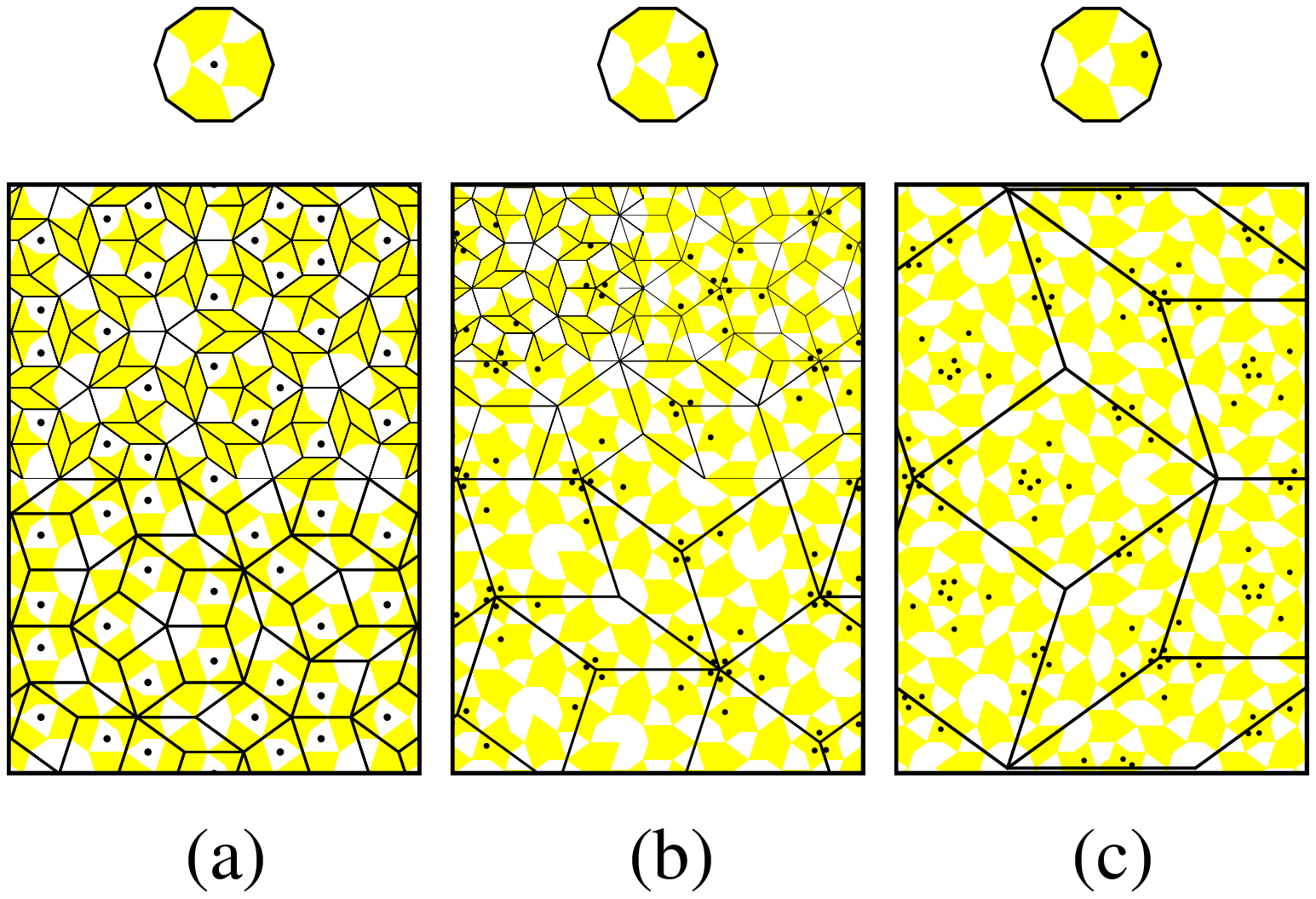} 
 \caption{
 }
 \label{f_atom_deco}
 \end{figure}

\begin{figure}[!b]
 \TwoFigPlace{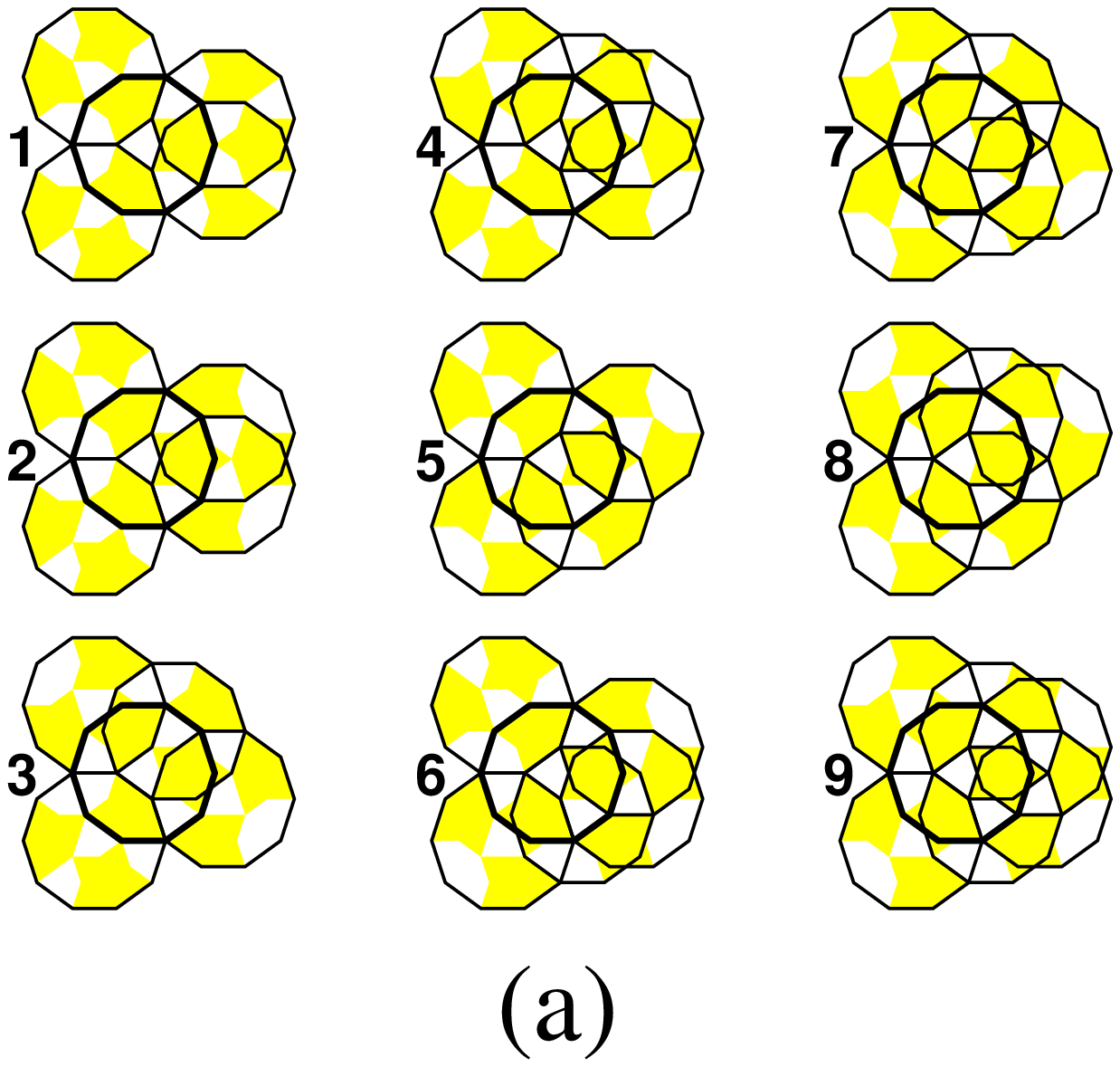}{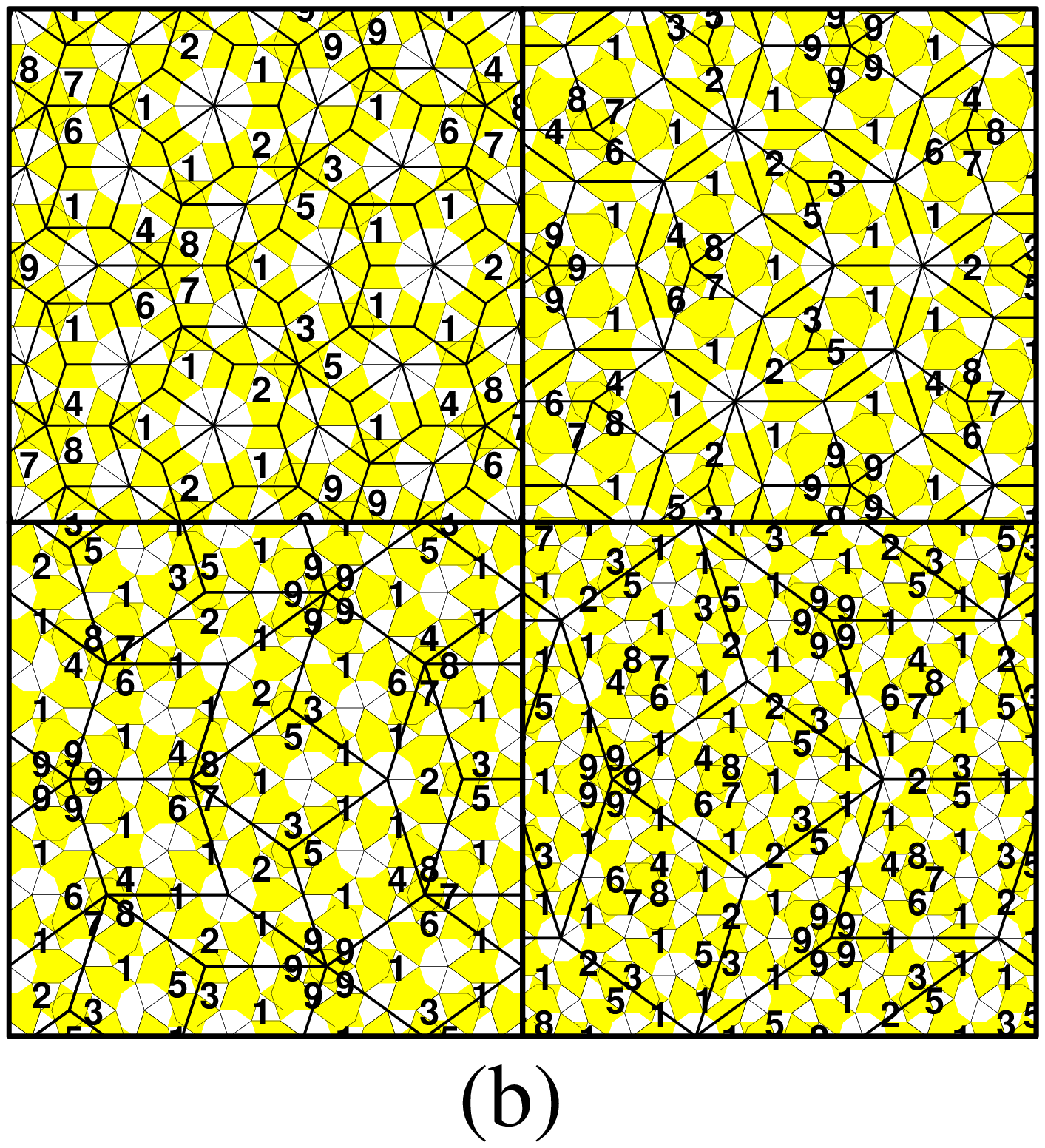}{0.}{65.}{65.}{-0.}{0.}{245.}{-50.}
 \caption{
 }
 \label{f_4th_inflated}
 \end{figure}

\begin{figure}[!h]
 \includegraphics[width=70mm]{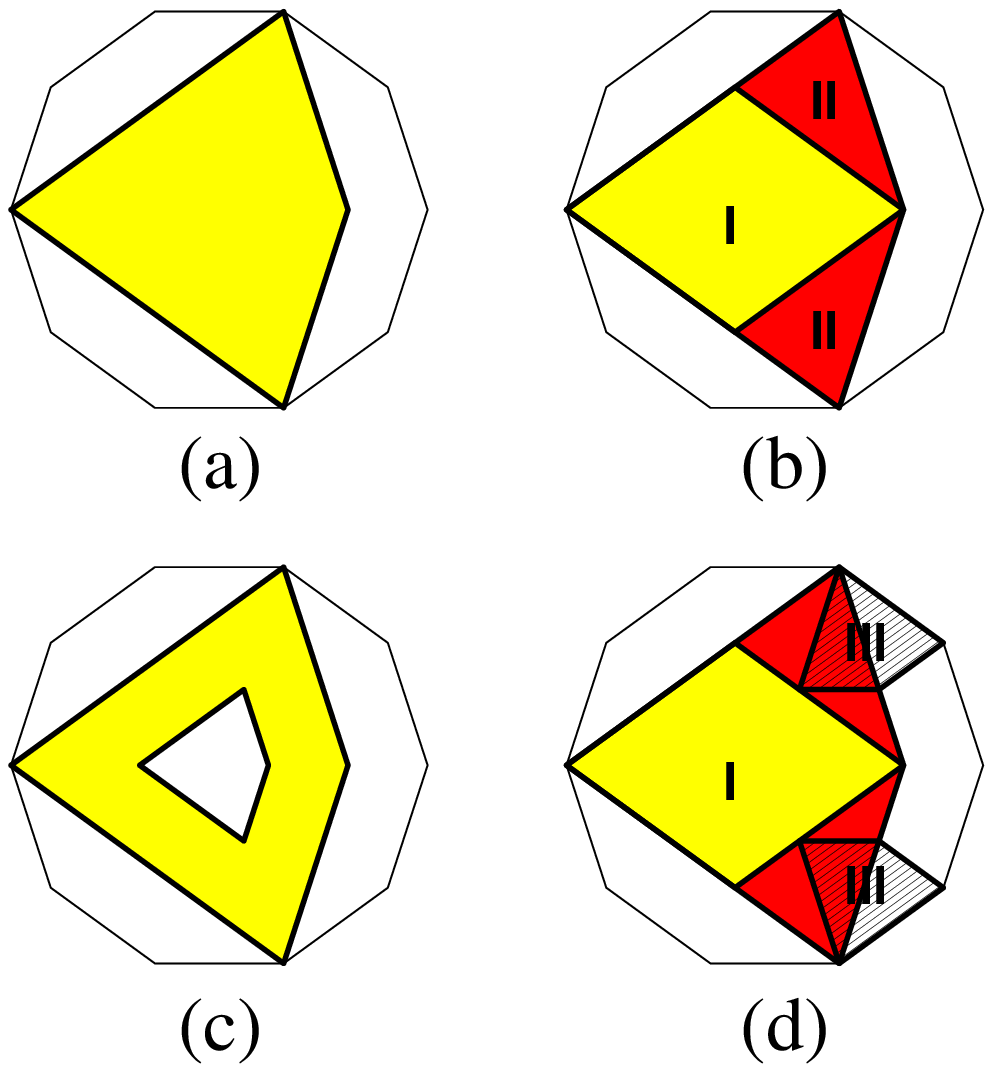} 
 \caption{
 }
 \label{f-region}
 \end{figure}

\begin{figure}[!b]
 \includegraphics[width=90mm]{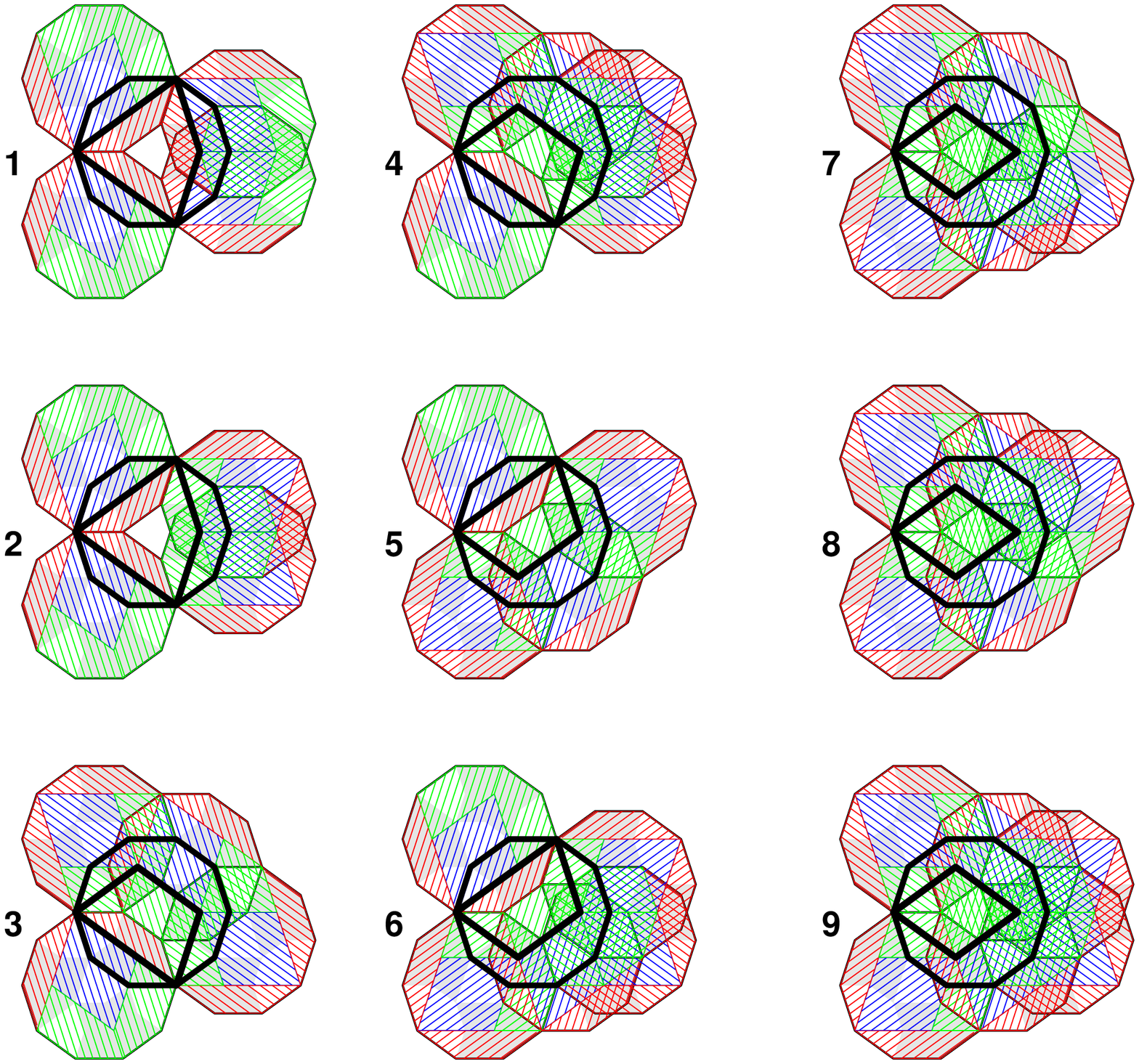} 
 \caption{
 }
 \label{f-9confOL}
 \end{figure}

\begin{figure}[!h]
 \includegraphics[width=90mm]{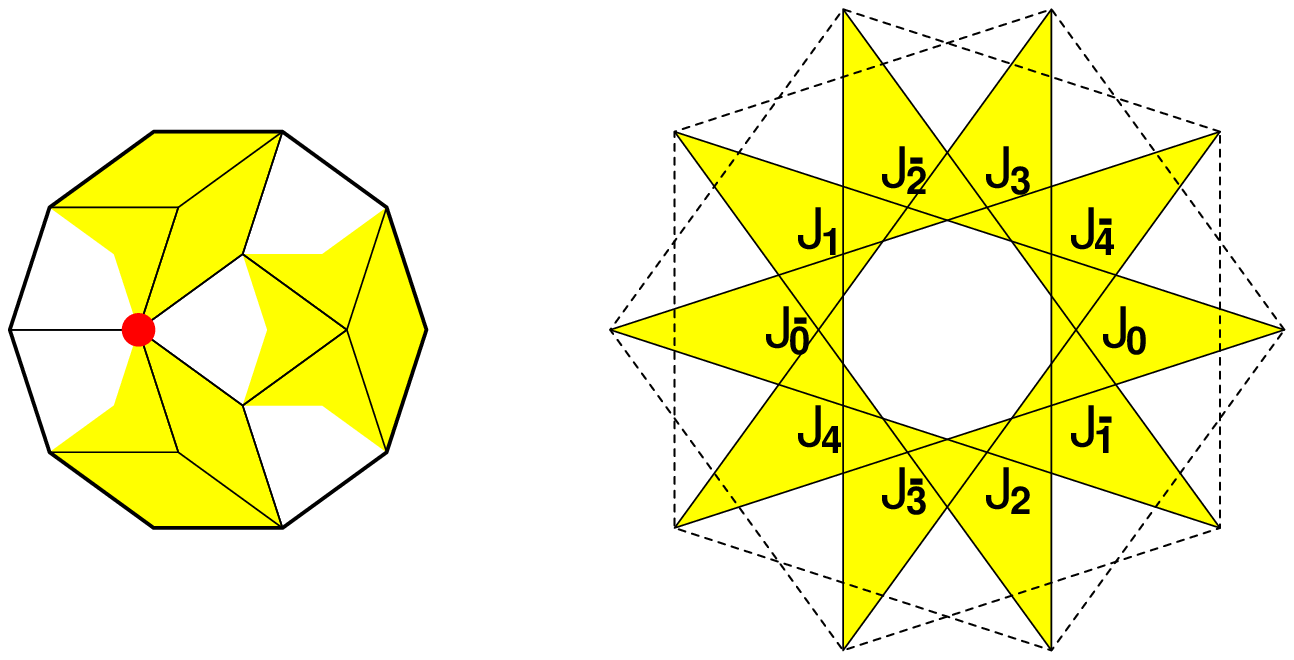} 
 \caption{
 }
 \label{f_deco_Jack}
 \end{figure}

\begin{figure}[!b]
 \TwoFigPlace{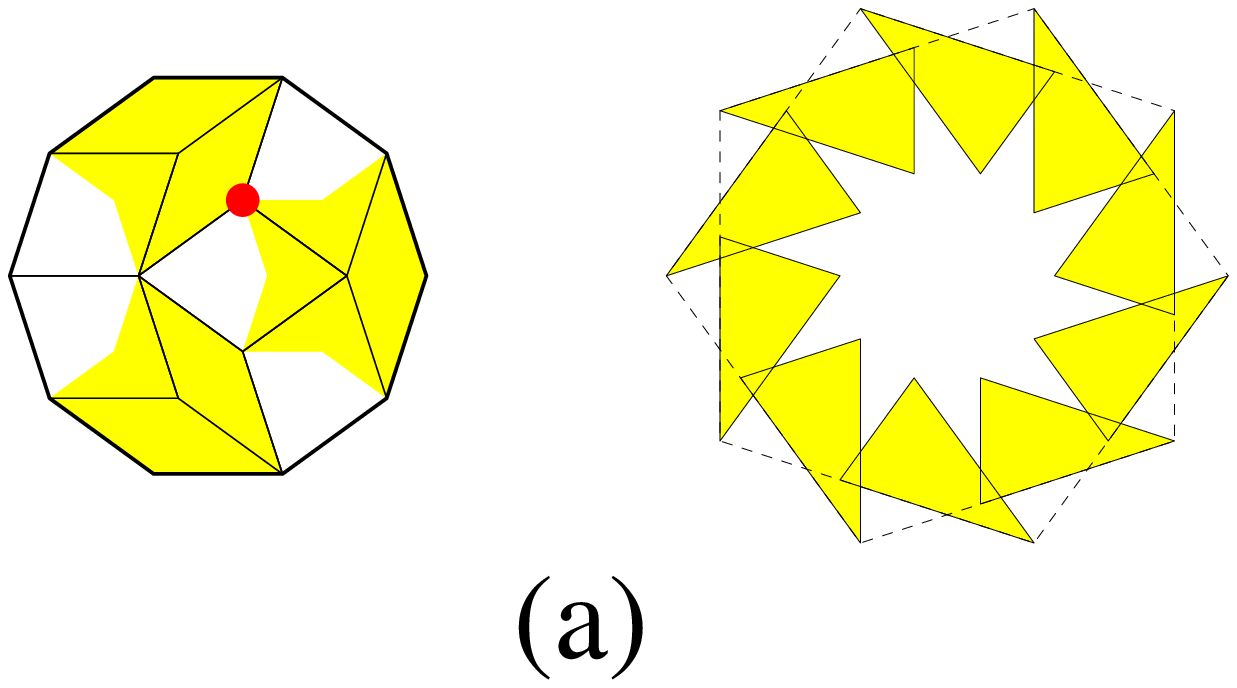}{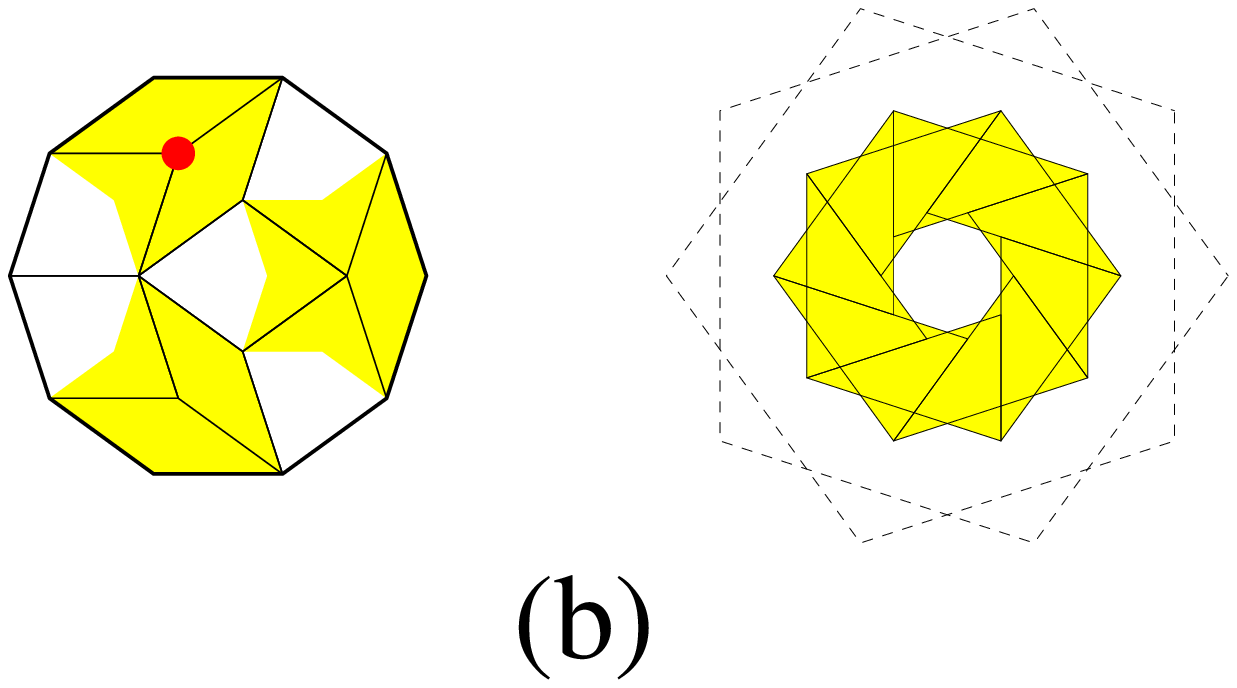}{0.}{70.}{70.}{10.}{10.}{90.}{-60.}
 \caption{
 }
 \label{f_deco_no_m}
 \end{figure}

 \begin{itemize}
 \item[{\bf Fig.~\ref{f_overlap_rule}:}] 
 Marked decagon and overlapping rules which force a quasiperiodic tiling. 
 (a) Marked decagon can force a quasiperiodic tiling
 when it is arranged by the Gummelt overlapping rules.
 (b) Gummelt overlapping rules demand that two decagons may overlap only 
 if shaded regions overlap and if the overlap are is no smaller than
 the hexagon overlap region in A.
 \item[{\bf Fig.~\ref{f_tiling}:}] 
 Decagon covering can be mapped into a conventional
 tiling by decorating decagon properly.
 (a) Rhombus Penrose tiling by dividing up central area of
 decagon into Jack.
 (b) Three-tile core-area tiling using the core area assignment
 of Jeong and Steinhardt~\cite{Jeong97}.
 (c) Original Penrose tiling~\cite{Penrose74} of 4-tile tiling
 by joining the centers of decagons.
 \item[{\bf Fig.~\ref{f_atom_deco}:}] 
 Quasi-unit-cell models and Penrose-tile models.
 (a) Decagon decoration at the top is not a decoration of original
 Penrose tile (upper part) but a decoration
 of single inflated tiles (lower part).
 (b) This decagon decoration is not a decoration of original (upper left)
 Penrose tile, nor single inflated (upper right) tile.
 Neither doubly (middle) nor triply (bottom) inflated Penrose tiles are 
 decorated identically.
 (c) Decagon decoration of (b) is a Penrose decoration of quadratically
 inflated Penrose tiles. 
 \item[{\bf Fig.~\ref{f_4th_inflated}:}] 
 (a) Nine ways of surrounding a decagon found in a Gummelt decagon
 covering~\cite{Gummelt96}.
 The numbers represent the surrounding configuration types.
 (b) Typed decagon arrangements in the decagon covering.
 The decagon arrangements in all inflated tiles of the 
 same shapes are identical but their types are not identical for doubly
 and triply inflated tiles. For the quadratically inflated 
 fat tiles, the types of decagons are also identical
 as shown in the right-lower panel.
 \item[{\bf Fig.~\ref{f-region}:}] 
 (a) The big kite for FOD. For an FOD model, image sites lying
in the shaded big kite region must be fundamentally decorated.
 (b) The regions for the number density weights for the FOD models. 
Weights $\rh{I} = \rh{0}$ and $\rh{II} = \tau^{-1} \rh{0}$ are
assigned to the atoms in the regions~I and~II respectively.
 (c) The big and the tiny kites for AOD. For an AOD model, image sites 
lying in the shaded region must be fundamentally decorated.
 (d) The regions for the number density weights for the AOD models. 
The region~II is the same as in (b). Weights for the region~I and~II
are the same as the FOD case. For atoms in the regions~III,
weight 0, $\rh{III}$, or $\rh{III}'$ can be assigned (see text for detail).
\item[{\bf Fig.~\ref{f-9confOL}:}] 
Overlaps between the core area of the central decagon and
the surrounding decagons. The region III in Fig.~\ref{f-region}(d)
can overlap with the tiny central kite area of the central 
decagon for Conf.-3.,\ldots,9.
 \item[{\bf Fig.~\ref{f_deco_Jack}:}] 
 A decoration of the fundamental decagon and the perp-space image of 
 the resulting quasi-unit-cell model.
 When the decagon covering is converted to the rhombus Penrose tiling
 by dividing up central area of decagon into a Jack, the decorated
 atom position corresponds to the Jack vertex.
 In the perp-space image, $J_{k}$ and $J_{\bar{k}}$ represent the 
 perp-space image of Jack 
 vertices in the $\hat{e}_k$ direction and the $-\hat{e}_k$ direction 
 decagons respectively. 
 \item[{\bf Fig.~\ref{f_deco_no_m}:}] 
 (a) A decoration of the fundamental decagon which results in a 
 quasi-unit-cell model with no mirror symmetry. 
 Both Fundamental decoration and the perp-space image have no
 mirror symmetry axis. 
 (b) An asymmetric decoration of the fundamental decagon
 which results in a quasi-unit-cell model with the mirror symmetry.
 Although the fundamental decagon is decorated asymmetric, decagons
 in the covering are decorated with the mirror symmetry due to the
 overlap and their perp-space image has the mirror symmetry.
 \end{itemize}

\clearpage

 \mbox{}

 \mbox{}

 \mbox{}

\clearpage

 \mbox{}

\clearpage
 \mbox{}

 \mbox{}

\clearpage

 \mbox{}

 \end{document}